\documentclass{llncs}
\usepackage{makeidx}  
\usepackage[utf8]{inputenc}
\usepackage{graphicx} 
\usepackage{blindtext}
\usepackage{chngcntr}

\usepackage{xargs} 
\usepackage[colorinlistoftodos,prependcaption,textsize=tiny]{todonotes}
\newcommandx{\comment}[2][1=]{\todo[linecolor=red,backgroundcolor=red!25,bordercolor=red,#1]{#2}}

\begin{document}
\frontmatter          
\pagestyle{headings}  
%
\mainmatter              
\title{Deliberative Platform Design: The case study of the online discussions in Decidim Barcelona}
\titlerunning{Deliberative Platform Design: The case study of the online discussions in Decidim Barcelona}  
%

\author{
Pablo Aragón\inst{1,2} \and 
Andreas Kaltenbrunner\inst{2} \and 
Antonio Calleja-López\inst{3} \and 
Andrés Pereira\inst{4} \and 
Arnau Monterde\inst{3} \and 
Xabier E. Barandiaran\inst{5,6} \and 
Vicenç Gómez\inst{1}
}
\authorrunning{Pablo Aragón et al.} 
%
\tocauthor{Pablo Aragón, Andreas Kaltenbrunner, Antonio Calleja-López, Andrés Pereira, Arnau Monterde, Xabier E. Barandiaran and Vicenç Gómez}
\institute{
Universitat Pompeu Fabra, Spain
\and
Eurecat - Technology Center of Catalonia, Spain
\and
Internet Interdisciplinary Institute - Universitat Oberta de Catalunya, Spain
\and
aLabs.org, Spain
\and
Ajuntament de Barcelona, Spain
\and
School of Social Work, UPV/EHU - University of the Basque Country, Spain
}



\maketitle              

\begin{abstract}
With the irruption of ICTs and the crisis of political representation, many online platforms have been developed with the aim of improving participatory democratic processes. However, regarding platforms for online petitioning, previous research has not found examples of how to effectively introduce discussions, a crucial feature to promote deliberation. In this study we focus on the case of Decidim Barcelona, the online participatory-democracy platform launched by the City Council of Barcelona in which proposals can be discussed with an interface that combines threaded discussions and comment alignment with the proposal. This innovative approach allows to examine whether neutral, positive or negative comments are more likely to generate discussion cascades. The results reveal that, with this interface, comments marked as negatively aligned with the proposal were more likely to engage users in online discussions and, therefore, helped to promote deliberative decision making.

\end{abstract}

\section{Introduction}

The crisis of representative democracy in the last three decades~\cite{rosanvallon2008counter,tormey2015end} has been identified with the crisis of democracy itself~\cite{della2013can,keane2009life}. Some authors have criticized the technocratic tendencies operating in this period as signs of the rise of post-democracy~\cite{crouch2004post} or post-politics~\cite{ranciere2001ten,vzivzek2000ticklish}, while others, more precisely, have used the term ``post-representation'', to refer to the emptying out (of power and meaning) of representative institutions by dynamics ranging from globalization to growing citizen mistrust~\cite{brito2008representation,keane2009life}. Specially in the last years, this political crisis has led to a period of fertile democratic innovation supported by an intensive and creative use of ICTs~\cite{castells2009communication,toret2013tecnopolitica}. Thus, we are witnessing new forms of participatory and deliberative democracy based on computer mediated communication~\cite{fuchs2007internet,hague1999digital}.

One of the recent institutional instantiations of this wider democratizing process is \emph{Decidim Barcelona}\footnote{\url{https://www.decidim.barcelona/}}, an online platform developed by the Barcelona City Council for supporting its participatory processes, e.g., the development of the Barcelona's strategic city plan. The strategic city plan defines objectives and actions to be carried out by the local government during the present legislature. The goal of this participatory process was to enroll the citizenry in a two month process of co-production, where citizens could discuss and support the proposals made by the government; and also make, discuss and support their own proposals. In total, more than 40~000 citizens participated in this process.

According to the functional specification of \emph{Decidim Barcelona}~\cite{monterde2015pam}, different pre-existing tools for participatory democracy were assessed, in particular, \emph{e-Petitions Gov UK} (United Kingdom)\footnote{\url{https://www.gov.uk/petition-government}}, \emph{Your Priorities} (Iceland)\footnote{\url{https://www.yrpri.org}}, \emph{Cónsul} (Madrid)\footnote{\url{https://decide.madrid.es/}}, and \emph{Open Irekia} (Basque Country)\footnote{\url{http://www.irekia.euskadi.eus/}}. On the one hand, these four tools share certain commonalities. First, they are web applications based on Free/Libre and Open Source Software (FLOSS). Second, they have been deployed in real environments by city, regional, or national governments. Third, they allow users to make online proposals. On the other hand, there are many differences among these four platforms. An important one is the way proposals are discussed by users. In \emph{e-Petitions Gov UK}, proposals cannot be discussed and, therefore, this tool might be considered as enabling participatory but not deliberative democracy. \emph{Your Priorities} allows users to publish comments either supporting the proposal (hereafter \emph{positive comments}) or against it (hereafter \emph{negative comments}). Positive and negative comments are displayed in two columns and sorted by the number of votes they receive to show the best arguments and, ultimately, to facilitate decision making. Although this strategy relies on comments, users do not engage in discussions, which might reduce the deliberative capabilities of the platform. In contrast, \emph{Cónsul} corresponds to an opposite scenario given that users are able to discuss any proposal with a threaded interface without any visual indication of whether comments are positive or negative. Finally, the approach in \emph{Open Irekia} allows users to indicate whether a comment is positive, negative or neutral. However, neutral, positive and negative comments are presented separately without applying a threaded discussion interface, as done in \emph{Cónsul}. 
This heterogeneity received special attention in the design specification process of \emph{Decidim Barcelona}~\cite{monterde2015pam} resulting in an interface which hybridizes the previous approaches. On the one hand, proposals are discussed in a threaded interface to promote online discussions and, consequently, online deliberation. On the other hand, users are able to establish when posting a first level comment (i.e., a direct comment to a proposal) whether is positive, negative or neutral in relation to the proposal. In addition, authors of proposals and comments are notified when receiving replies.

Figure~\ref{fig:interface} (see appendix) shows a real proposal for the strategic city plan which requested a municipal ice skating rink in Barcelona. The discussion page shows two first positive (green) comments with no replies and a third negative (red) comment calling into doubt the adequacy of expending public funding on a winter sport facility in a Mediterranean city. As shown in Figure~\ref{fig:interface}, the negative comment triggered a discussion cascade among users. 

This proposal is an illustrative example of the aim of this hybrid interface: users can engage in online discussion to promote deliberative processes while positive and negative comments are easily distinguishable to facilitate decision making. The combination of both approaches makes \emph{Decidim Barcelona} an interesting case study for multiple reasons. First, recent studies have shown that conversation threading in online discussion platforms promotes the emergence of discussion cascades with higher levels of reciprocity~\cite{aragon2017ICWSM} and online deliberation~\cite{aragon2017detecting}. Second, given that users are able to mark the alignment of comments with the proposal (positive, negative and neutral), we can compare the typical network structures originated by the different types of comment alignment. According to~\cite{GonzalezBailonJIT2010}, these structures can be used as proxies of very basic forms of deliberation.
Given this particular scenario of \emph{Decidim Barcelona}, the research question of this study is as follows:
\begin{itemize}
\item \emph{Which are the structural differences of discussion cascades triggered by neutral, positive or negative comments on online proposals?}
\end{itemize}

As presented in the following section, despite the increasing research work on online petition plaforms, how to effectively introduce discussions is an open practical and research challenge~\cite{lindner2009electronic}. We postulate that the combination in \emph{Decidim Barcelona} of both conversation threading and comment alignment (in particular, explicitly negative comments to the proposal) should favor cognitive dissonance~\cite{festinger1962theory} in users, which would lead to a higher willingness to discuss the proposals, and, therefore, to  deliberative practices of decision making.

The organization of the paper is as follows. In Section~\ref{sec:relatedwork} we present related work on online petition platforms. We then describe in Section~\ref{sec:dataset} our dataset of discussion threads from \emph{Decidim Barcelona}. Next, we introduce the structural metrics of discussion threads for our analysis in Section~\ref{sec:metrics} and the results of the study in Section~\ref{sec:analysis}. Finally, in Section~\ref{sec:discussion}, we discuss the implications of our findings for the design of online participation platforms.

\section{Related work}\label{sec:relatedwork}
The interest in online petition platforms is reflected by the increasing attention from academia~\cite{panagiotopoulos2012online}. Some of the first studies analysed the platform developed by the German Parliament either to identify different types of users according to the frequency of participation~\cite{jungherr2010political} or to characterize the relationship between online participation and offline socio-demographic factors~\cite{lindner2011broadening}.  
Indeed, much effort has been made to detect which factors affect the signing of online petitions~\cite{anduiza2008online,Hale2013PGS,huang2015activists,margetts2015leadership,yasseri2013modeling}.

Previous work has also examined the impact of platform design on the dynamics of online petitioning. A study of the UK government petitions platform showed that introducing trending information on the homepage increased the inequality in the number of signatures across petitions~\cite{hale2014investigating}. In relation to our research question, some papers have precisely assessed the role of the availability and design of discussion features. A study of online petition platforms launched by UK local authorities (Kingston and Bristol)~\cite{whyte2005petitioning} examined the performance of the online forums incorporated in these tools. Results indicated that most users did not visually identify the possibility to discuss proposals  and just a few users published comments. Therefore, the study concluded that the discussion section for online petitions needed to be more appealing. A comparative analysis of four online petition systems (the aforementioned platform of the German Parliament and the platforms of the Scottish Parliament,
the Parliament of Queensland,
and Norwegian municipalities)
also examined whether they integrate an online discussion forum~\cite{lindner2009electronic}. Online discussions were available in every platform except for the case of Queensland. The study found little usage of these forums and concluded with the open research question about the function of these discussions and how to channel them into the political decision-making processes.

\section{Dataset}\label{sec:dataset}
Our dataset contains the discussion threads from the proposals in \emph{Decidim Barcelona} for the development of the strategic city plan. To better understand the discussions that originated more activity we present in Table~\ref{tab:top} (see Appendix) the most commented proposals, which are related to controversial topics in Barcelona like housing affordability and mobility.

Data were extracted through the Decidim API\footnote{\url{https://www.decidim.barcelona/api/docs}} to obtain a total of 10~860 proposals and 18~192 comments. 16~217 comments were first level comments (i.e., direct replies to the proposal) while 1~975 comments were replies to comments. As mentioned in the introduction, users were able to establish the alignment of first level comments with the proposal. Thus, 10~221 comments were marked as neutral (63.03\%), 5~198 comments were marked as positive (32.05\%), and only 798 comments were marked as negative (4.92\%).

\section{Structural metrics of discussion cascades}\label{sec:metrics}
\begin{figure}[ht!]
\centering
\includegraphics[trim={0cm 0cm 0cm 0cm},clip, angle=100, width=0.4\textwidth]{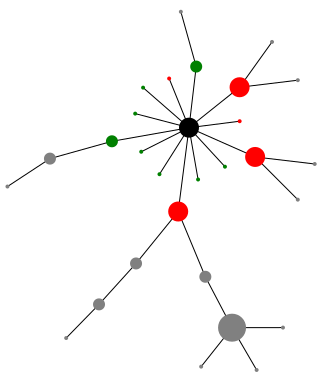}
\caption{Radial tree visualization of the proposal presented in Figure~\ref{fig:interface}. Black node (root) represents the proposal, green nodes are positive and red nodes negative first level comments. Comment nodes are sized by the indegree (number of replies to the comment). The visualization shows a cascade of comments triggered by a negative comment to the proposal (red node on the right).} \label{fig:tree}
\end{figure}

Discussion threads are collections of messages posted as replies to either an initial message (the proposal) or another message (a comment). For this reason, discussion threads can be represented as a directed rooted tree. We present in Figure~\ref{fig:tree} the proposal for a municipal ice skating rink (shown in Figure~\ref{fig:interface}) using a radial tree visualization tool~\cite{aragon2016ICWSMdemo}. The black node is the proposal (root) and the nodes directly connected to the root are the first level comments (green colored if positive and  red colored if negative). This tree structure allows to identify whether a first level comment triggers a discussion cascade, e.g., the red node on the right, which is the negative comment against expending public funding on a winter sport facility in a Mediterranean city, 
triggers several comments. 

The structure of the discussion cascade of each first level comment can be characterized with typical metrics of tree graphs:
\begin{itemize}
\item size: number of nodes,
\item width: maximum number of nodes at any level,
\item depth: number of levels,
\item h-index: maximum level $h$ in which there are, at least, $h$ comments~\cite{gomez2008statistical}.
\end{itemize}
In the discussion cascade originated by the aforementioned negative comment about public funding (red node on the right in Figure~\ref{fig:tree}), size is 9, width is 4, depth is 3, and h-index is 3. 

With the exception of the size, which just quantifies the volume of the cascade, these metrics serve to inform about the network topology of a cascade. Morever, the last three metrics have been suggested to quantify the level of deliberation in online discussion threads~\cite{GonzalezBailonJIT2010}. This approach is based on the Madisonian conceptualization of deliberation as the conjugation of two dimensions: representation and argumentation~\cite{ackerman2002deliberation}. Given that messages at any level often represent users within the discussion, width has been proposed to quantify the extent of representation of the online community in a discussion cascade. Because the exchange of arguments between users commonly occur as exchange of comments, the depth of the discussion cascade (i.e., the largest exchange of comments) has been proposed to capture argumentation. The last structural metric (h-index) both considers width and depth and, therefore, has been proposed to measure online deliberation in a discussion cascade~\cite{GonzalezBailonJIT2010}.

\section{Analysis of discussion cascades}\label{sec:analysis}

\begin{figure}[!b]
\centering
\includegraphics[width=0.65\textwidth,trim={3cm 9cm 4cm 10cm},clip]{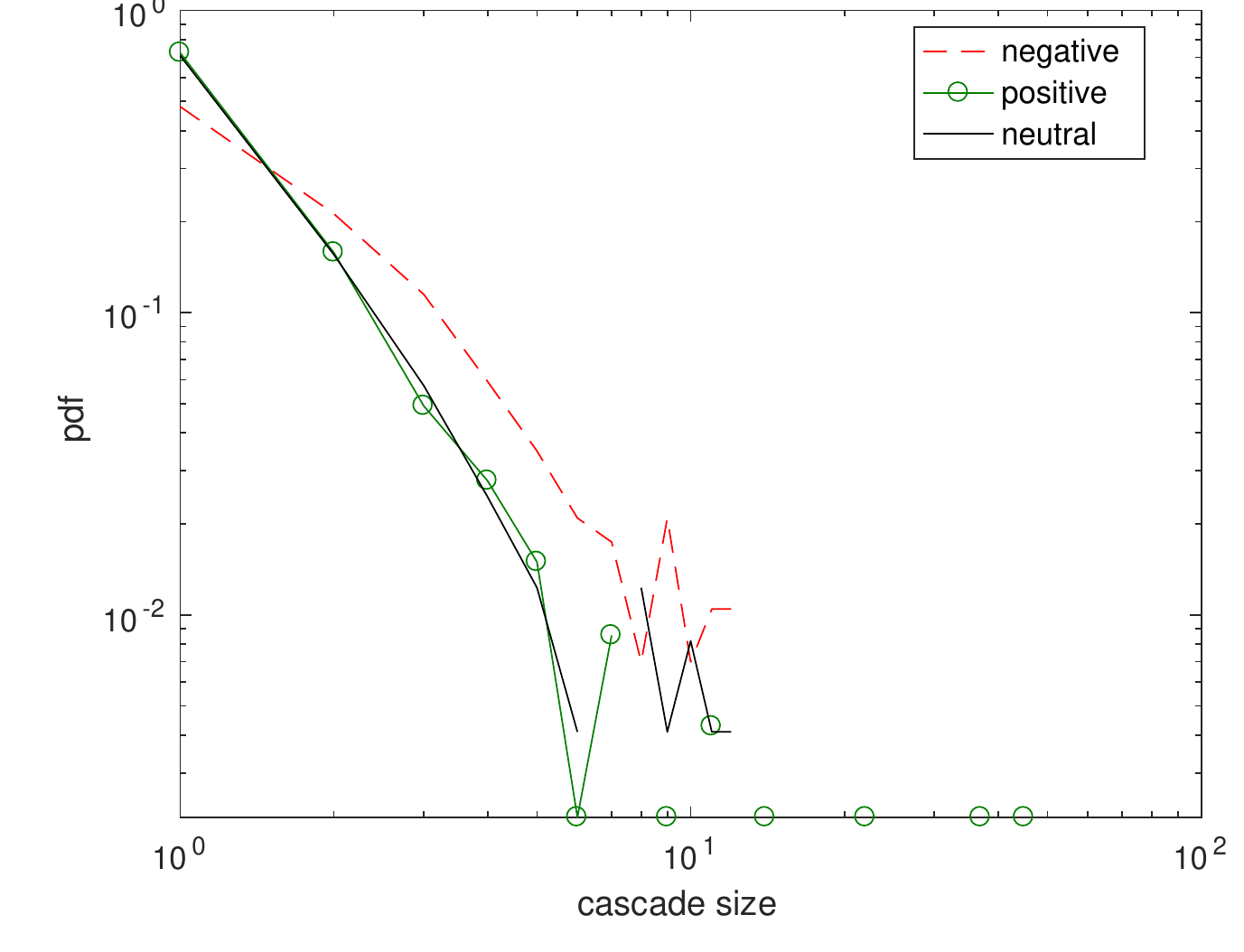}
\caption{Distribution of the cascade size triggered by the first level comments of each alignment (neutral, positive and negative).} \label{fig:size-alignment}
\end{figure}

The description of the dataset indicated that most of the first level comments were marked as neutral, an important fraction were marked as positive and just around 5\% were marked as negative. To understand the structure of cascades triggered by comments from different alignments, we first examine the  distribution of the cascade size depicted in Figure~\ref{fig:size-alignment}. We observe a notably higher probability of triggering a cascade for negative comments. We also observe that, for every alignment, few cascades contain more than five comments.

\begin{figure}[ht]
\centering
\includegraphics[width=1\textwidth]{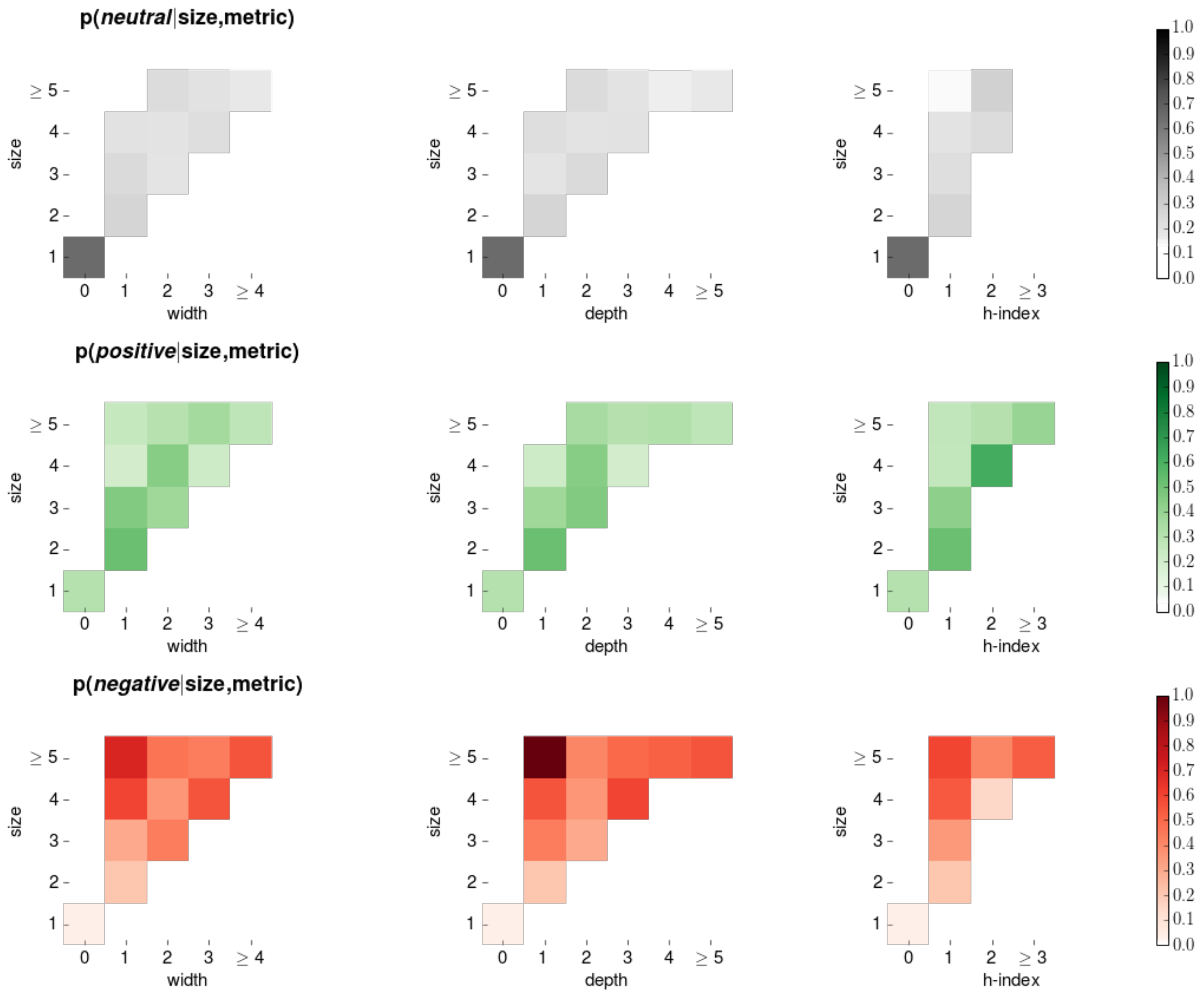}
\caption{Heatmaps of the probability of alignment (gray for neutral, green for positive and red for negative) of a first level comment given size and width, depth, or h-index of the cascade. Large values are aggregated in the top rows and rightmost columns.} \label{fig:alignment-size-metric}
\end{figure}

Figure~\ref{fig:size-alignment} reveals as well a larger preference for larger cascades triggered by negative comments. However, the size of the cascade is not an informative metric of the structure of the cascade. For this reason, we examine the probability of the alignment of the root (comment) of the cascade with different sizes and different values of the structural metrics (width, depth and h-index). Results are presented in Figure~\ref{fig:alignment-size-metric} using heatmaps, i.e., the darker the more likely. We observe that, if a comment did not trigger any discussion cascade, that comment is probably neutral or positive. In contrast, when comments originated discussion, there is a  higher probability that they are negative. Furthermore, the likelihood of negative comments increases when the value of the size and the structural metrics also increase.

\begin{figure}[ht]
\centering
\includegraphics[width=1\textwidth]{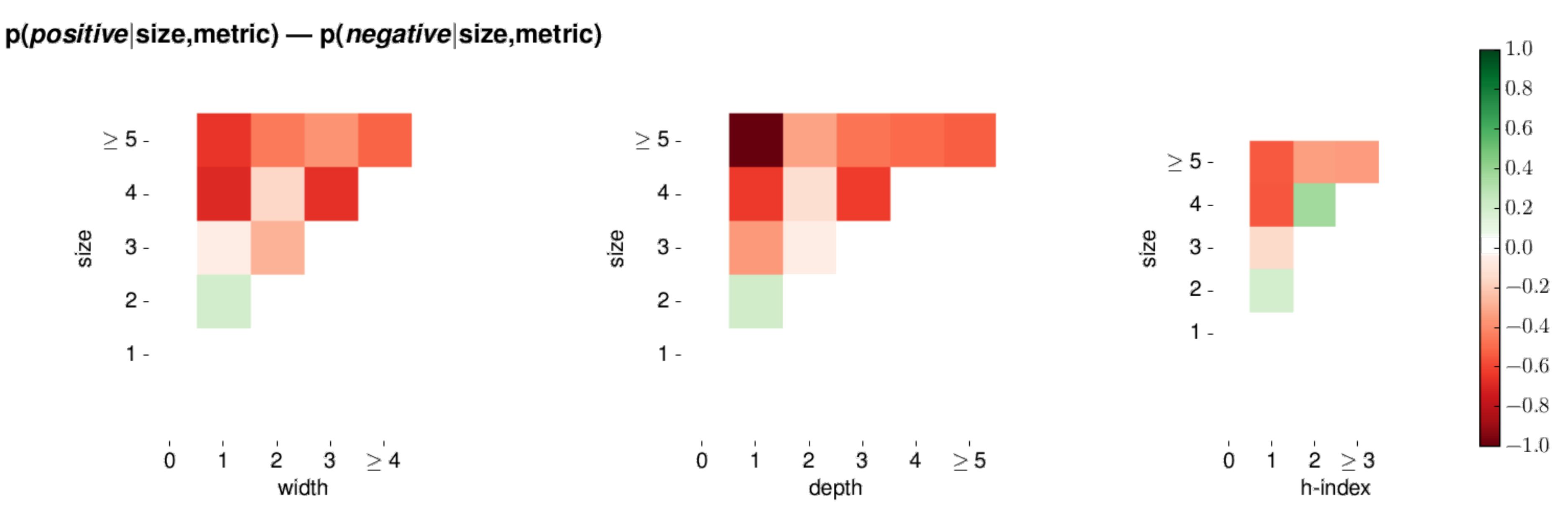}
\caption{Heatmaps of the probability of polar alignment (green for positive and red for negative) of a first level comment given the value of size and structural feature (width, depth, and h-index) of the cascade. Values are obtained with a statistical test of 10K evaluations with 10K random cascades each and shown if significant ($p<0.05$).} \label{fig:comparative}
\end{figure}

These results suggest that discussion cascades occur more frequently due to negative messages and less frequently due to neutral messages. However, to perform a rigorous analysis we need to consider the following observations. First, we found by manual inspection that many neutral comments, despite being clearly positively or negatively aligned with the proposal, were not explicitly marked accordingly for some reason, e.g., problems of usability or perhaps a deliberate choice of the user. Second, we have to take into account the class imbalance (5~198 positive vs. 798 negative comments). Because of these two reasons, we will restrict our analysis to aligned comments, either positive or negative,
which triggered at least one reply. We apply bootstrapping, with 10K evaluations and randomly chosen (with replacement) 10K positive and 10K negative comments. Comments can be chosen more than once. The number of evaluations and threads have been selected, after multiple assessments, to guarantee the significance of the statistical test ($p<0.05$). Results are presented as heatmaps in Figure~\ref{fig:comparative} and confirm that, regarding positive and negative first level comments, when deep and complex cascades are observed, there is a much stronger likelihood to be originated by a negative comment. In conclusion, although we find that positive comments sometimes triggered complex discussion cascades, in general, the deepest and most complex conversations between users in \emph{Decidim Barcelona} were caused by negative comments, i.e., counter-argumentation.

\section{Discussion}\label{sec:discussion}
This study has been designed to answer our research question about the structural differences of discussion cascades triggered by neutral, positive and negative comments on online proposals in~\emph{Decidim Barcelona}. Our question was motivated by the
open research challenge of effectively deploying online discussions in online petition platforms~\cite{lindner2009electronic,whyte2005petitioning}. The interface in \emph{Decidim Barcelona}, which combines conversation threading and comment alignment, became an innovative case study and an ideal scenario to answer this question. Results are clear: although a low proportion of comments were negative (about 5\%), negative comments were more likely to trigger more complex discussion cascades than neutral and positive comments. We should note that users in \emph{Decidim Barcelona} were notified when they received a reply. Therefore, authors of proposals were always aware of negative comments which might also increase their interest in engaging in discussion to advocate for their proposals. This is consistent with the basis of cognitive dissonance~\cite{festinger1962theory}, i.e., negative comments usually contain new information which contradicts the idea of a given proposal and the author and supporters of the proposal will be likely to reply to it.  We can conclude, thus, when trying to address the open challenge of effectively combining online petitioning and online discussion~\cite{lindner2009electronic,whyte2005petitioning}, the deliberative platform design of \emph{Decidim Barcelona} introduces an innovative solution.

We should remark that our methodology was language-independent. This was a deliberated decision because of the complexity of the bilingual context of \emph{Decidim Barcelona} (Spanish and Catalan), e.g., many natural language processing resources were not available for Catalan. Although this decision allows to easily apply our methodology on any other platform, future work should also focus on the content of messages to compare how linguistic features might also differ in relation to the alignment of comments.

\section*{Acknowledgments}
This work is supported by the Spanish Ministry of Economy and Competitiveness under the María de Maeztu Units of Excellence Programme (MDM-2015-0502).

\bibliographystyle{splncs03}
\bibliography{references}

\newpage\appendix\section{Supporting Information}
\counterwithin{figure}{section}
\counterwithin{table}{section}
\setcounter{figure}{0}    

\begin{figure}[ht]
\centering
\includegraphics[width=0.99\linewidth,trim={0cm 0cm 0cm 0cm}, clip]{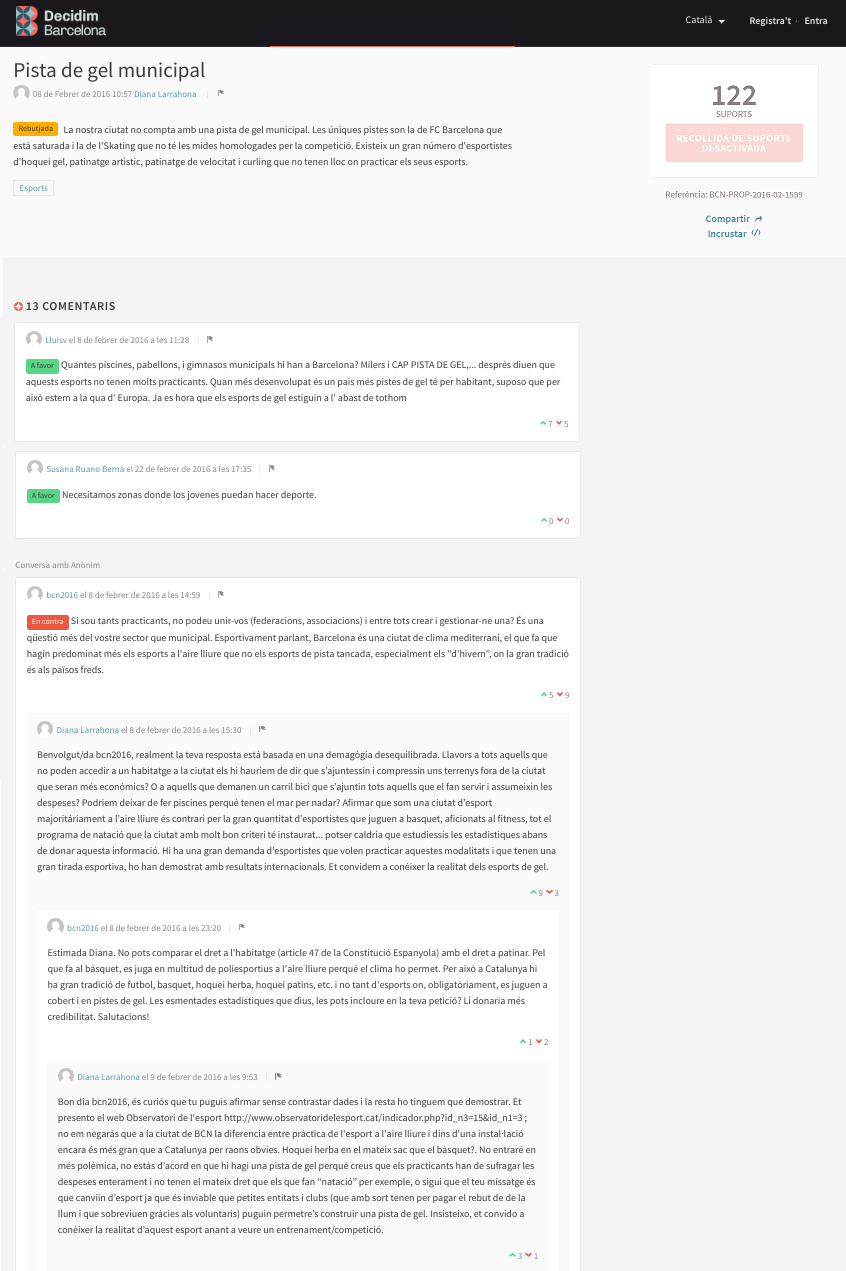}
\caption{Discussion page of a proposal in \emph{Decidim Barcelona} for building a municipal ice skating rink. The hybrid interface combines both conversation threading and coloring, i.e., positive and negative first level comments include green and red labels, respectively (the interface at that time colored the full text of positive and negative comments).} \label{fig:interface}
\end{figure}

\begin{table}[ht]
\centering
\caption{Top proposals in \emph{Decidim Barcelona} by the number of comments. An English translation is indicated in parentheses.}
\label{tab:top}
\begin{tabular}{lr}
\textbf{Title}                                                                                                                                                                     & \textbf{N. comments} \\\hline
\begin{tabular}[c]{@{}l@{}}Noves llicències per a pisos turístics\\ (New licenses for tourist apartments)\end{tabular}                                                                 & 337               \\\hline
\begin{tabular}[c]{@{}l@{}}Implantar el tramvia a la Diagonal\\ (To build a tramway in Diagonal Avenue)\end{tabular}                                                                                 & 111               \\\hline
\begin{tabular}[c]{@{}l@{}}Cubriment de la Ronda de Dalt al seu pas per la Vall d'Hebrón\\ (Roof for Dalt Road in Vall d'Hebrón)\end{tabular}                                      & 108               \\\hline
\begin{tabular}[c]{@{}l@{}}Promoció de l'ús de la bicicleta, i millora i ampliació dels carrils bici\\ (Promotion of cycling and improvement and expansion of bike lanes)\end{tabular} & 80                \\\hline
\begin{tabular}[c]{@{}l@{}}Regulació del mercat de lloguer\\ (Regulation of the housing rental market)\end{tabular}                                                                            & 77               
\end{tabular}
\end{table}

\end{document}